\documentclass[runningheads]{llncs}

\usepackage[linesnumbered]{algorithm2e}
\usepackage{longtable}
\usepackage{color}

\usepackage{subfig}
\usepackage{caption}

\usepackage{graphicx}

\usepackage{float}

\usepackage{amsmath,epsfig,} 

\usepackage{hyperref}
\usepackage{cleveref}
\usepackage{rotating}
\usepackage{comment}

\usepackage{hyperref}

\newcommand{\footlabel}[2]{%
    \addtocounter{footnote}{1}%
    \footnotetext[\thefootnote]{%
        \addtocounter{footnote}{-1}%
        \refstepcounter{footnote}\label{#1}%
        #2%
    }%
    $^{\ref{#1}}$%
}

\usepackage{xcolor}
\definecolor{boxgrey}{HTML}{F3F3F3}

\newlength\mylen
\newcommand\myinput[1]{%
  \settowidth\mylen{\KwIn{}}%
  \setlength\hangindent{\mylen}%
  \hspace*{\mylen}#1\\}
  
\newcommand{\hlbox}[2]{%
  \setlength{\fboxsep}{5pt}
  \begin{center}%
    \fcolorbox{white}{boxgrey}{%
      \parbox{.95\columnwidth}{\noindent \textbf{#1}. \textit{#2}}
    }%
  \end{center}%
}

\usepackage{amssymb} 

\usepackage{lineno}
\modulolinenumbers[5]

\AtBeginDocument{%
  \providecommand\BibTeX{{%
    \normalfont B\kern-0.5em{\scshape i\kern-0.25em b}\kern-0.8em\TeX}}}

\begin{document}

\title{\textsc{MOReGIn}: Multi-Objective Recommendation\\ at the Global and Individual Levels}

\titlerunning{MOReGIn: Multi-Objective Recommendation}

\author{Elizabeth G\'omez\inst{1}\orcidID{0000-0003-2698-3984}
\and
David Contreras\inst{2}\orcidID{0000-0002-6906-485X}
\and
Ludovico Boratto \inst{3}\orcidID{0000-0002-6053-3015} 
\and
Maria Salam\'o \inst{1}\orcidID{0000-0003-1939-8963}
}

\authorrunning{G\'omez et al.}

\institute{Facultat de Matem\`atiques i Inform\`atica, Universitat de Barcelona, Barcelona, Spain \\\email{egomezye13@alumnes.ub.edu,maria.salamo@ub.edu} \and
Facultad de Ingeniería y arquitectura, Universidad Arturo Prat, Iquique, Chile \email{david.contreras@unap.cl} \and Department of Mathematics and Computer Science, University of Cagliari, Cagliari, Italy \email{ludovico.boratto@acm.org}}

\maketitle

\begin{abstract}
Multi-Objective Recommender Systems (MORSs) emerged as a paradigm to guarantee multiple (often conflicting) goals. 
Besides accuracy, a MORS can operate at the {\em global} level, where additional beyond-accuracy goals are met for the system as a whole, or at the {\em individual} level, meaning that the recommendations are tailored to the needs of each user. 
The state-of-the-art MORSs either operate at the global or individual level, without assuming the co-existence of the two perspectives.
In this study, we show that when global and individual objectives co-exist, MORSs are not able to meet both types of goals.
To overcome this issue, we present an approach that regulates the recommendation lists so as to guarantee both global and individual perspectives, while preserving its effectiveness. 
Specifically, as individual perspective, we tackle genre calibration and, as global perspective, provider fairness.  
We validate our approach on two real-world datasets, publicly released with this paper\footref{footLabel}.
\end{abstract}


\section{Introduction}
{\bf Motivation.} Since the goal of recommender systems is to provide relevant suggestions for the users, the main focus has been the effectiveness of the results~\cite{RicciRS22}.
Nevertheless, users might be interested in properties of the items besides their effectiveness, and there are other stakeholders who can benefit from how recommendations are produced (e.g., content providers). 
Hence, beyond-accuracy perspectives are central to the generation and evaluation of recommendations.

{\em Multi-Objective Recommender Systems} (MORSs) support the provision of perspectives that go beyond item relevance, such as, e.g., diversity, calibration, and fairness~\cite{ZhengW22}. The optimization for these objectives can happen at the {\em global} ({\em aggregate}) level, thus ensuring that the system as a whole can guarantee certain properties (e.g., all providers receive a certain exposure in the recommendation lists). 
In alternative, a MORS can operate at the {\em individual} ({\em local}) level, and shape results that are consider the prominence of individual users towards the different goals (e.g., each user can receive a different level of diversity or the recommended genres can be calibrated to the preferences in the training set)~\cite{Jannach22}.

When analyzing the current literature, a MORS either operates at the global level~\cite{GeZYPHHZ22,LiCFGZ21,LinCPSXSZOJ19,NaghiaeiRD22,wu2022multifr} or at the local level~\cite{PeskaD22,DokoupilPB23,DokoupilPB23b}.

\vspace{2mm}
\noindent {\bf Open issues.} There might be scenarios in which both global and individual objectives co-exist. 
Indeed, a platform might decide that, as a whole, the recommendations should offer certain properties (e.g., be fair to providers of different demographic groups, or enable a certain level of novelty). 
Moreover, specific goals might be set for the individual users (e.g., the calibration of the genres or the diversity of the recommended items might need to follow what is observed in the training set of each user). 
As we show in Section~\ref{sec:assessment}, when a MORS tackles {\em only} global or individual perspectives, the other perspective trivially remains under-considered and cannot be guaranteed by the system.

\vspace{1mm}
\noindent {\bf Our contributions.} To overcome the aforementioned challenges, in this paper, we present a MORS that produces recommendations with both global and individual objectives. 
As a use case, we consider, as a global objective, {\em provider fairness} and, as an individual one, {\em calibrated recommendations}. 
This aligns our study with the rest of the MORS literature, where two beyond-accuracy objectives are considered. 
For the sake of clarity, we will talk about {\em provider-fair and calibrated recommendations} but, as we discuss in Section~\ref{subsec:metrics}, {\bf our approach can be generalized to any global or individual objective}.

Besides accounting for beyond-accuracy perspectives involving both global and individual objectives, the problem of providing provider-fair and calibrated recommendations becomes interesting also from a practical point of view. As we will show in Section~\ref{sec:exploratory}, users tend to rate items of certain genres and that are produced in certain geographic areas, suggesting that we can account for both perspectives at the same time when generating the recommendations. Hence, at the technical level, we would need a unique solution that (i) produces effective results for the users, (ii) can provide fairness for providers belonging to different groups at the {\em global} level, i.e., by distributing, over the entire user base, the recommendation of items belonging to different provider groups in equitable ways, and (iii) can calibrate the recommendation lists of each {\em individual} user. 

Our approach involves a post-processing strategy. 
To enable a form of provider fairness that can consider demographic groups that are not necessarily characterized by a binary group (e.g., males and females), we consider, as a sensitive attribute, the geographic provenance of the providers and have the different continents as the granularity with which we split the groups; this is aligned with recent literature on provider fairness~\cite{GomezBS22,GomezZBSR22}. 
As in classic calibrated fashion, we distribute the recommendations according to the item genre.
Based on this characterization of the data, we present an approach that makes use of buckets to associate the continents in which the items are produced and the genre of the items. 
We use these buckets to post-process the recommendation lists (we will later discuss that this is the best way to regulate both aggregate- and individual-level properties) and regulate how the recommendations are distributed across the users. Thanks to the fact that each bucket contains (i) the continent in which the item is produced, to regulate provider fairness, and (ii) the genre of the item, to regulate calibration, both global and individual perspectives are captured at once by our approach.
To validate our proposal, we apply it to the recommendations produced by five algorithms, and study the effectiveness of our approach on two datasets (including a novel one, released with this study), and against state-of-the-art approaches for calibrated recommendation and provider fairness.

Concretely, our contributions can be summarized as follows:
\begin{itemize}
    \item After the identification of the research gaps (Section~\ref{sec:related}) and characterization of our setting (Section~\ref{sec:preliminaries}), we provide the foundations to our use case, by showing that calibration and provider fairness are related problems, since the genres of the items and their country of production are connected (Section~\ref{sec:exploratory});
    \item We present an approach to post-process the recommendation lists to meet both global and individual goals. We calibrate the results for the individual users in terms of genres, and are fair towards providers (Section~\ref{sec:approach});
    \item We face the limitation of evaluating this problem, due to the scarcity of data offering both the category of the items and the sensitive attributes of the providers, so we i) extend the MoviLens-1M dataset, to integrate the continent of production of each item, and ii) we collect and present (in Section~\ref{sec:preliminaries}) a novel dataset. Both resources are publicly available \href{https://www.dropbox.com/sh/9hbvshlox0qbhi9/AADlAoaOwCQF1yLCjLpJb-CVa?dl=0}{here}\footlabel{footLabel}{https://tinyurl.com/yc6nnx5v};
    \item We perform experiments (Section~\ref{sec:assessment}) to validate our proposal when applied to the recommendation produced by five algorithms, covering both memory- and model-based approaches, and point-wise and pair-wise approaches. To evaluate its effectiveness in different domains, we consider movie and song recommendation as application scenarios. Based on our outcomes, we highlight possible research paths that might emerge from it (Section~\ref{sec:conclusions}).
\end{itemize}

\section{Related Work}\label{sec:related}

\vspace{1mm}
\noindent{\bf MORSs.} Recent literature has studied how to account for multi-objective goals from different angles. The user perspective was tackled by Li \textit{et al} \cite{LiCFGZ21}, which balance recommendation accuracy for users with different levels of activities. From an item perspective, Ge \textit{et al.} \cite{GeZYPHHZ22} proposed an approach to balance item relevance and exposure. Considering both the user and item perspectives, Naghiaei \textit{et al.} \cite{NaghiaeiRD22} propose a re-ranking approach to account for  consumer and provider fairness. Other studies blend the multiple objectives into a single function, in order to obtain a Pareto-optimal solution~\cite{LinCPSXSZOJ19,wu2022multifr}. Recent advances have also proposed MORSs in sequential settings, by optimizing the results for accuracy, diversity, and novelty~\cite{StamenkovicKAXK22}.
MORS that operate at the individual level have optimized the recommendation process mainly via online interactions, such as conversational approaches \cite{10.5555/3327546.3327641} or via critiquing \cite{WANG2020106369,elahiinteractive}, but approaches aiming at learning individual propensities from past interactions also exist, e.g., \cite{JUGOVAC2017321,PeskaD22,DokoupilPB23,DokoupilPB23b}.

\vspace{1mm}
\noindent{\bf Calibrated recommendation.} {\em Calibration} is a well-studied technique commonly used to solve the problem of unfair output~\cite{ref_nixon2019measuring,ref_zadrozny2001obtaining,ref_calibration} in recommender systems. 
Seymen \textit{et al.} \cite{ref_seymen2021constrained} address the problem of providing calibration in the recommendations from a constrained optimization perspective. Abdollahpouri \textit{et al.} \cite{ref_abdollahpouri2020connection} 
study the connection between popularity bias, calibration, and consumer fairness in recommendation. Recently, Rojas \textit{et al.} \cite{ref_Rojas} analyze how the calibration method in~\cite{ref_calibration} deals with the bias in different recommendation models. Other studies focus on analyzing user profiles to mitigate miscalibrated recommendations~\cite{ref_lin2020calibration} or to mitigate popularity bias from the user's perspective~\cite{DBLP:journals/apin/ChenWSZH23}.   
Existing metrics have some limitations when applying a user-centered approach to evaluate popularity bias and calibrated recommendations. To address these limitations, 
 Abdollahpouri \textit{et al.} \cite{ref_abdollahpouri2021user} present a new metric. 

\vspace{1mm}
{\bf Provider fairness.} {\em Provider fairness} ~\cite{Ekstrand0B022}   
has been studied in many common scenarios, e.g.,~\cite{MarrasBRF21,GomezZBSR22,EkstrandTKMK18,EkstrandKluverUMUAI:2021,GharahighehiVP:2021,FerraroSB21,MehrotraMBL018}. 
It is usually assessed by considering metrics such as the visibility and the exposure that respectively assess the amount of times an item is present in the rankings~\cite{FabbriBB020,ZehlikeB0HMB17} and {\em where} an item is ranked~\cite{BiegaGW18,Zehlike020}, for users to whom each provider’s items are recommended. 
Other approaches, such as that by Karakolis \textit{et al.} \cite{app12104984}, consider 
diversity and coverage for users. 
Raj \textit{et al.} \cite{DBLP:conf/sigir/RajE22} present a comparative analysis among several fairness metrics recently introduced to measure fair ranking. Wu \textit{et al.} \cite{10.1145/3477495.3532007}
formalize a family of exposure fairness metrics that model the problem of fairness jointly from the perspective of both types of stakeholders.

\vspace{1mm}
{\bf Contextualizing our work.} No MORS can address both calibrated recommendation lists for the users and provider fairness. Our algorithm's aims are
 to provide i) each user with calibrated recommendations, ii) fair recommendations for the providers, iii) aiming at a minimum loss in effectiveness.

\section{Preliminaries}\label{sec:preliminaries}

\subsection{Recommendation Scenario}\label{subsec:recommendationscenario}
Let $U = \{u_1, u_2, ..., u_n\}$ be a set of users, $I = \{i_1, i_2, ..., i_j\}$  be a set of items, and $V$ be a totally ordered set of values that can be used to express a preference together with a special symbol $\bot$. The set of ratings results from a map $r: U \times I \to V$, where $V$ is the rating domain. If $r(u,i)=\bot$, then we say that $u$ did not rate $i$. 
To easy notation, we denote $r(u,i)$ by $r_{ui}$. We define the set of ratings as $R=\{(u,i,r_{ui})\,:\,u\in U,\,i\in I,\, r_{ui}\neq\bot\}$ and they can directly feed an algorithm in the form of triplets (point-wise approaches) or shape user-item observations (pair-wise approaches). We denote with $R_u$ the ratings associated with a user $u \in U$. We consider a temporal split of the data, where a fixed percentage of the ratings of the users (ordered by timestamp) goes to the training and the rest goes to the test set~\cite{BelloginCC17}.
The goal is to learn a function $f$ that estimates the relevance ($\hat{r}_{ui}$) of the user-item pairs that do not appear in the training data (i.e., $r_{ui}=\bot$). We denote as $\hat{R}$ the set of recommendations.

Let $C$ denote the set of geographic continents in which items are organized. We consider a geographic continent as the provenance of an item provider. We denote as $C_i$ the set of geographic continents associated with an item $i$. Note that, since an item could be produced by more than one provider, it might be associated with several geographic continents, and thus, $|C_i|\geq1$ and $C_i \subseteq C$. In case two providers belong to the same geographic continent, that continent appears only once; indeed, we are dealing with group fairness so, when a group of providers is associated with an item (once or multiple times), we account for its presence.
We use the geographic continents to shape  demographic groups, which can be defined to group the ratings of the items produced in a continent (we denote the items in $I$ produced in a continent $c\in C$ as $I_c$, where $I_c \subseteq I$ ). 

Let $G$ denote the set of genres in which items are organized. We denote as $G_i$ the set of genres associated with an item $i$. Note that, an item can be of one or more genres, and thus, $|G_i|\geq1$ and $G_i \subseteq G$ . We denote the items in $I$ that have a genre $g\in G$ as $I_g$, where $I_g \subseteq I$. 

\subsection{Metrics}\label{subsec:metrics}

\vspace{1mm}
\noindent{\bf Provider-group Representation.} In order to enable provider fairness, we should understand the attention received by a provider group in the training data. For this reason, we compute the representation of a demographic group in the data as the number of ratings for items associated with that group in the data. We define with $\mathcal{R}$ the {\em representation} of a group $c\in C$ as follows:

\begin{equation}
\mathcal{R}_c =  \frac{\left|\{r_{ui}\,:\,u\in U,\, i \in I_c\}\right|}{|R|}
\label{eq:rep_cont}
\end{equation}

Eq.~\eqref{eq:rep_cont} accounts for the proportion of ratings given to the items of a demographic group associated with a continent. This metric ranges between 0 and 1. 
We compute the representation of a group only considering the training set.
Trivially, the sum of the representations of all groups is equal to 1.

\vspace{1mm}
\noindent{\bf User-based genre propensity.} In order to calibrate the results for the users, we need to understand how the preferences for the different item genres are distributed. For this reason, we define with $\mathcal{P}$ the {\em propensity} of a user of  $u\in U$ to rate items of a genre $g \in G$, as follows:

\begin{equation}
\mathcal{P}_{ug} = \frac{\left|\{r_{ui}\,:\,g\in G_i\}\right|}{|R_u|}
\label{eq:propensity}
\end{equation}

Eq.~\eqref{eq:propensity} accounts for the proportion of ratings associated with a genre for a given user. This metric ranges between 0 and 1. 
Trivially, the sum of the propensities of all genres for a user is equal to 1.
This metric is equivalent to the distribution $p(g|u)$~\cite{ref_calibration}.

\vspace{1mm} 
\noindent{\bf Disparate Impact.} We assess unfairness with the notion of {\em disparate impact} generated by a recommender system. Specifically, we assess disparate visibility.  

\begin{definition}[Disparate visibility] Given a group $c\in C$, the {\em disparate visibility} returned by a recommender system for that group is measured as the difference between the share of recommendations for items of that group and the representation of that group in the input data:

\begin{equation}
\Delta \mathcal{V}_c = \left(\frac{1}{|U|}\sum_{u \in U}\frac{|\{\hat{r}_{ui}: i\in I_c\}|}{|\hat{R}|}\right) - \mathcal{R}_c \label{eq:visibility}
\end{equation}

\end{definition}

\noindent The range of values for this score is $[-\mathcal{R}_c,1-\mathcal{R}_c]$; specifically, it is 0 when the recommender system has no disparate visibility, while negative/positive values indicate that the group received a share of recommendations that is lower/higher than its representation. This metric is based on that defined by Fabbri \textit{et al.} \cite{FabbriBB020}. 

\vspace{1mm} 
\noindent{\bf Miscalibration.} We assess the tendency of a system to recommend a user items whose genres are distributed differently from those they prefer via {\em miscalibration}.

\begin{definition}[Miscalibration] Given a user $u \in U$ and a genre $g \in G$, the miscalibration returned by a recommender system for that user is measured as the difference between the share of recommendations for items of that genre and
the propensity of the user for that genre in the training data:

\begin{equation}
\Delta \mathcal{M}_{ug} = \frac{|\{\hat{r}_{ui}: i\in I_g\}|}{|\hat{R_u}|} - \mathcal{P}_{ug} \label{eq:miscalibration}
\end{equation}
\end{definition}

\vspace{1mm} 
\noindent{\bf Generalizability.} The rest of our paper will consider disparate visibility ($\Delta \mathcal{V}_c$) as the {\em global} perspective and miscalibration ($\Delta \mathcal{M}_{ug}$) as the {\em individual} perspective our MORS considers. Nevertheless, our approach can be generalized to {\em any} metric that assesses the difference between (i) the distribution of the recommendations and (ii) what can be observed in the training set or an objective set by the platform via a policy (e.g., a given amount of content novelty or diversity).


\section{Matching Item Providers and Genre Propensity}\label{sec:exploratory}

\subsection{Real-world Datasets}\label{subsec:datasets}
First, we extended the MovieLens-1M dataset, so as to integrate the continent of production of each movie. Second, a domain that fits our problem is song recommendation. 
However, existing music datasets, such as LastFM-2B~\cite{MelchiorreRPBLS21}, do not contain song genres and sensitive attributes of the artists, so they do not fit our problem. Thus, we collected a dataset from an online music platform. 

In particular, the \textbf{MovieLens-1M (Movies)} dataset comprises 1M ratings (range 1-5), from 6,040 users for 3,600 movies across 18 genres. The dataset provides its IMDB ID, which allowed us to associate it to its continent of production, thanks to the OMDB APIs (\url{http://www.omdbapi.com/}). Keep in mind that {\em a movie may be produced on more than one continent}. On the other hand, \textbf{BeyondSongs (Songs)} contains 1,777,981 ratings (range 1-5), provided by 30,759 users, to 16,380 songs. For each song, we collected the continent of provenance of the artist, and 14 music genres. Both resources are available \href{https://www.dropbox.com/sh/9hbvshlox0qbhi9/AADlAoaOwCQF1yLCjLpJb-CVa?dl=0}{online}\footref{footLabel}.

\subsection{Characterizing Group Representation and Genre Propensity}\label{subsec:characterization} 

We consider the temporal split of the data, where 80\% of the ratings are considered for the training set and have been used to measure $\mathcal{R}_c$ and $\mathcal{P}_{ug}$.  
Note that, while the representation of a demographic group covers the entire training set, the propensity is measured at the user level. Hence, to characterize the link between the two phenomena we aggregate the propensity of all the users for a given genre by summing their values.

\begin{figure}[t]
\centering
\subfloat[Movies $\mathcal{R}_c$ \label{fig:movies_rep}]{
  \includegraphics[width=0.38\textwidth]{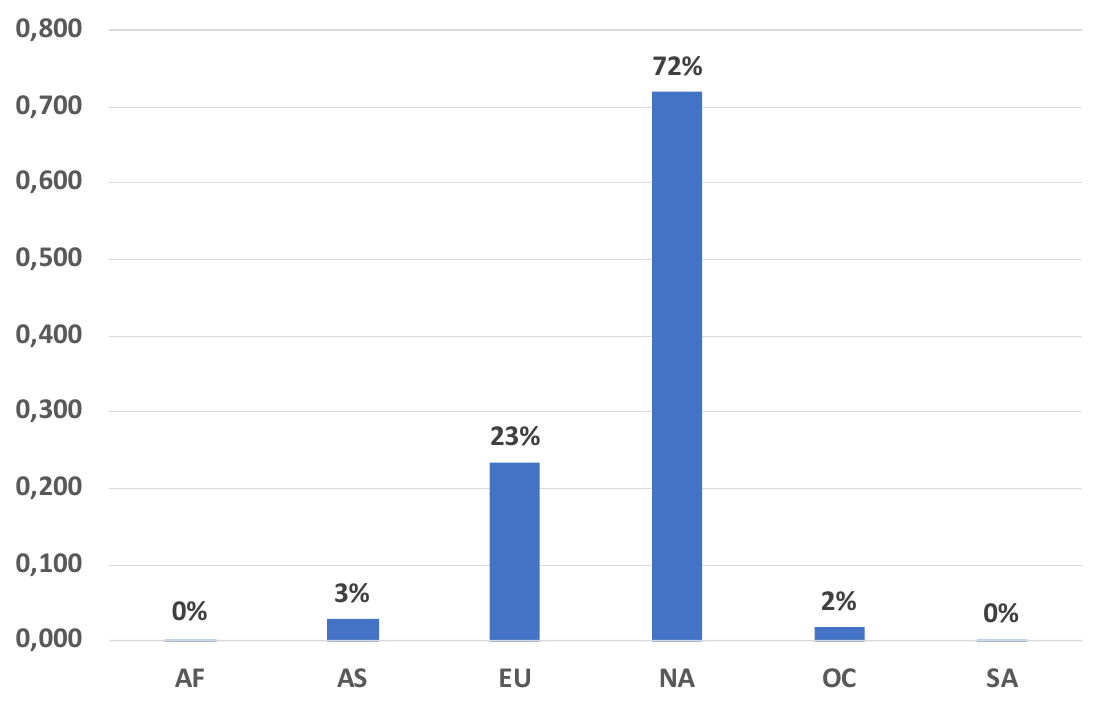}%
}\hfil
\subfloat[Songs $\mathcal{R}_c$ \label{fig:songs_rep}]{
  \includegraphics[width=0.38\textwidth]{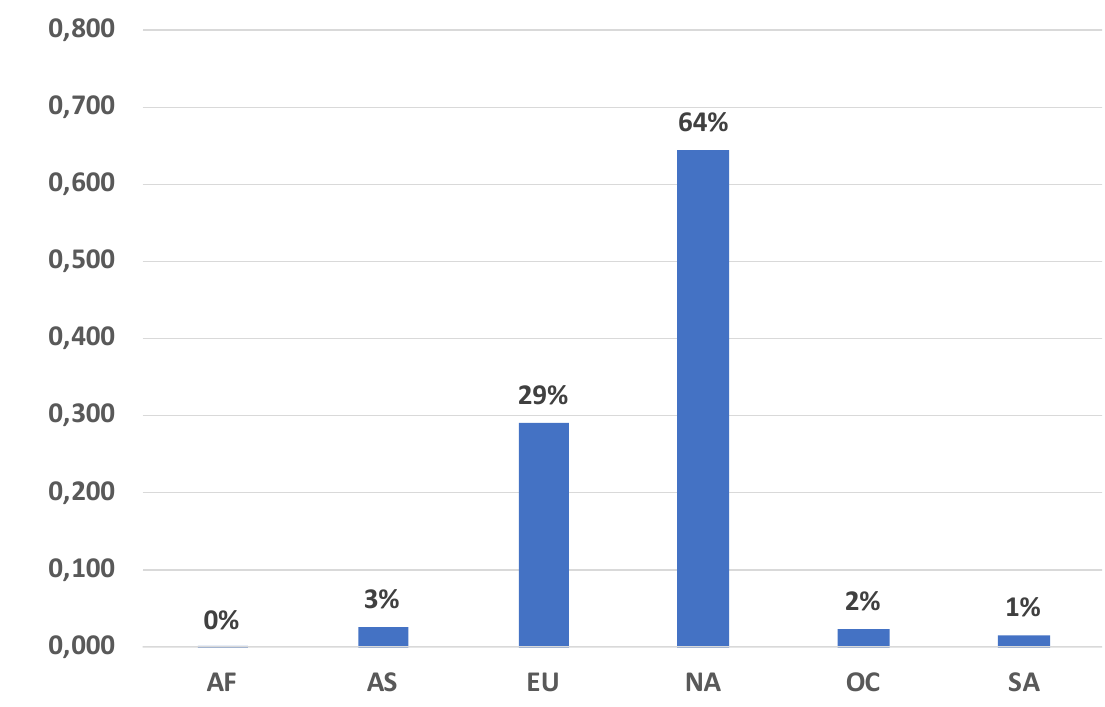}%
}\hfil
\vspace{-0.3cm}
\subfloat[Movies $\mathcal{P}_{ug}$ \label{fig:movies_prop}]{
  \includegraphics[width=0.48\textwidth]{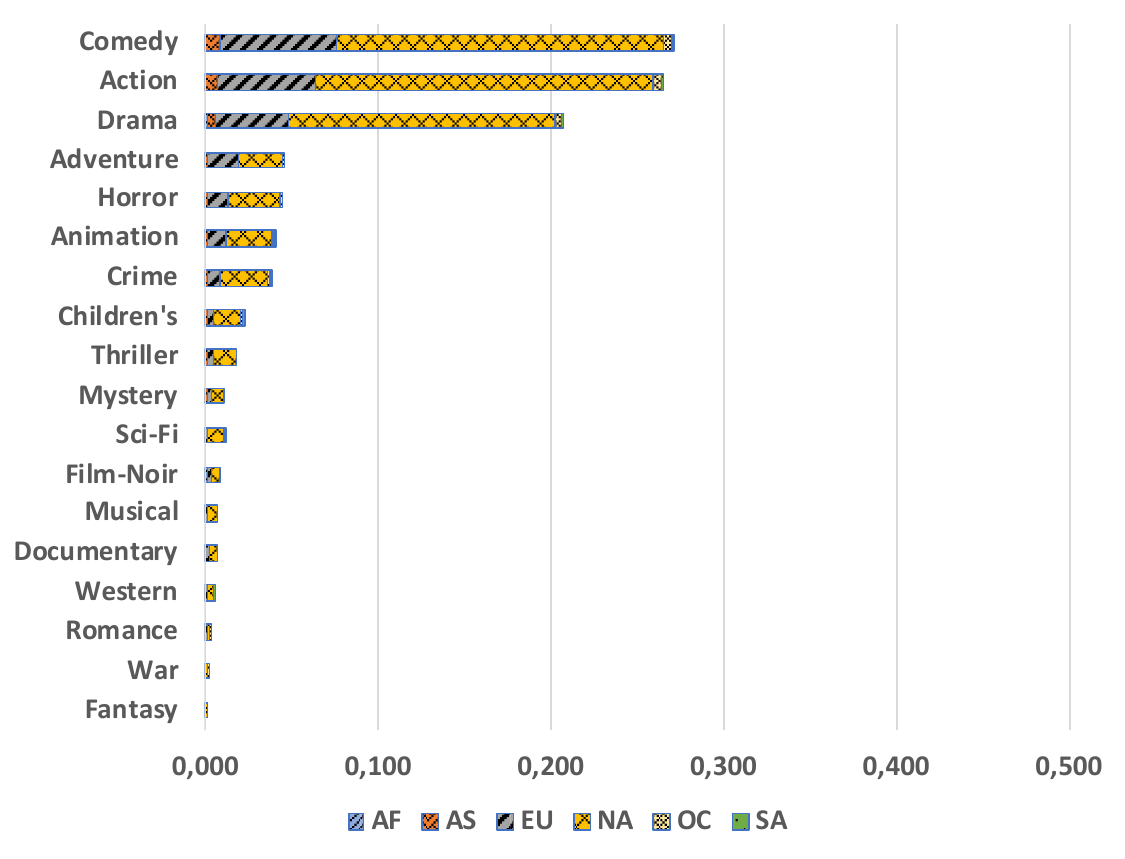}%
}\hfil
\subfloat[Songs $\mathcal{P}_{ug}$\label{fig:songs_prop}]{
  \includegraphics[width=0.48\textwidth]{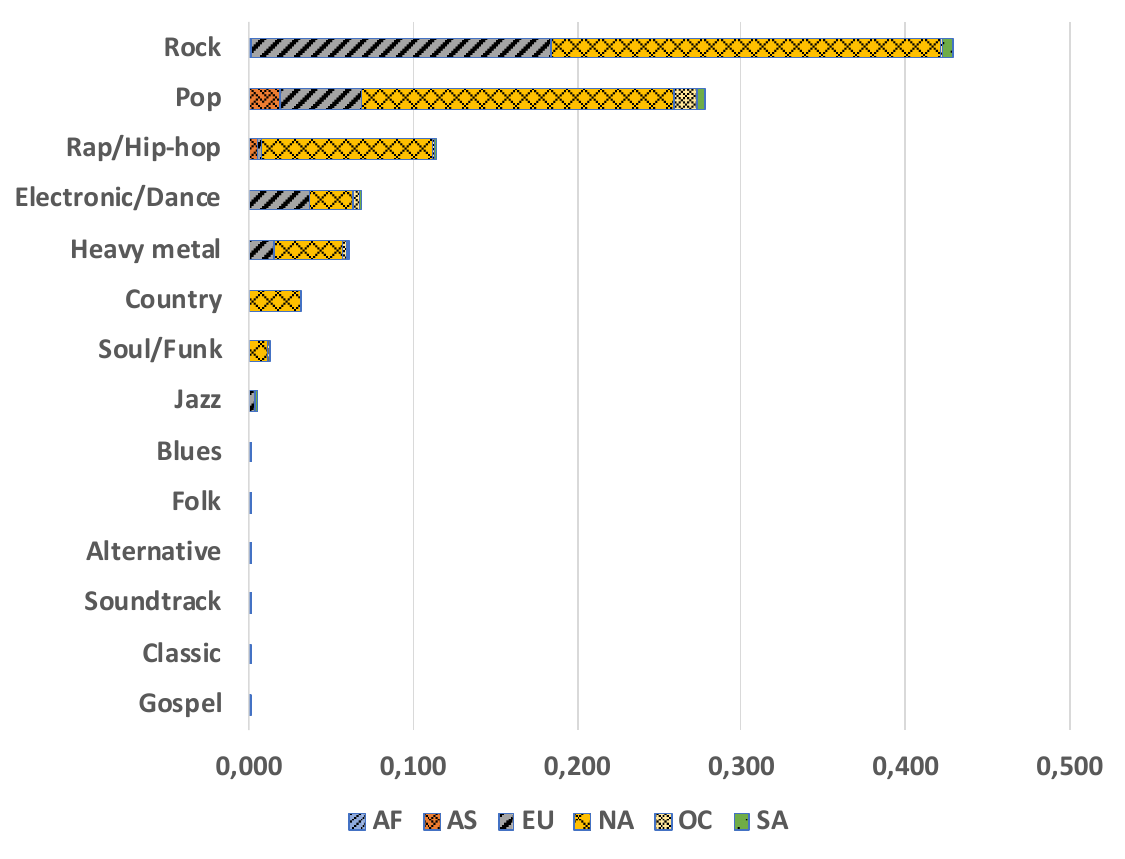}%
}
\caption{{\bf Group representation (a and b) and genre propensity (c and d) in the Movies and Songs data.} Acronyms stand for AF: Africa, AS: Asia, EU: Europe, NA: North America, OC: Oceania, SA: South America.} 
\vspace{-0.3cm}
\label{fig:MoviesSongsRepProp}
\end{figure}

Figures \ref{fig:movies_rep} and \ref{fig:songs_rep} show the $\mathcal{R}_c$ for Movies and Songs, respectively. Both datasets depict a similar representation by continents, where the highest representation is of items from NA providers (72\% in movies and 64\% in songs) and the second place is for EU providers (23\% and 29\%). 
In the rest of the continents, for both datasets, it is less than 10\%. Figures \ref{fig:movies_prop} and \ref{fig:songs_prop} show the $\mathcal{P}_{ug}$ in both datasets; three genres attract most of the ratings by users. 

We can also observe that ratings seem to be clustered between certain genre-continent pairs. In other words, different genres are distributed differently across continents. In the Movies data (\figurename~\ref{fig:movies_prop}), Comedy movies are largely preferred when produced by EU producers, just as Action attracted the majority of ratings for movies by NA producers. In the Songs data (\figurename~\ref{fig:songs_prop}), the Electronic/Dance genre was consumed much more heavily when produced by EU artists than by those in the rest of the world, and Heavy metal songs are mostly consumed when they come from NA. In both datasets, users' preferences for the minority provider groups (AF, AS, OC, and SA) are also concentrated on a few selected genres, confirming this rating aggregation in certain genre-continent pairs. 

\hlbox{Observation 1}{Users have the propensity to rate items of certain genres and that are produced by certain geographic groups (i.e., in certain continents). Calibration and provider fairness are related problems so, when producing recommendation lists, both perspectives should be accounted at the same time, in MORS fashion.}


\section{Individually Calibrated and P-Fair Recommendation}\label{sec:approach}

\subsection{Algorithm}\label{sec:algorithm}

\begin{algorithm}[t]
{ \footnotesize 
 \KwIn{$recList$: ranked list (records contain $user$, $item$, $rating$, $position$, $genre$ , $continent$), which arrives sorted by user and rating and contains $topn$ recommendations to the user. \\ 
 \myinput{$trainList$: list with the training set (records contain $user$, $item$, $rating$, $genre$, $continent$), which is sorted by user and rating.} \\ 
 \myinput{$topk$: top $k$ recommendations, we set up $k=10$.} \\
 \myinput{$topn$: top $n$ recommendations, we set up $n$= 1000.}
 }
 \KwOut{$reRankedList$: ranked list with Individually Calibrated and P-Fair Recommendation.} 
 \SetAlgoLined 
%
define \textbf{MOReGIn} ($recList$, $trainList$, $topk$, $topn$) \\
 \Begin{
    \tcp*[h]{\textcolor{blue}{Step 1. Compute $R_c$}}

    $recBucketRep$   $\leftarrow$ \textbf{computeRepresentation}($topk$, $recList$, $trainList$); 
         
    \tcp*[h]{\textcolor{blue}{Step 2. Compute $\mathcal{P}_{ug}$}}
       
    $recBucketUserProp$ $\leftarrow$ \textbf{computePropensity}($topk$, $recList$, $trainList$);

    \tcp*[h]{\textcolor{blue}{Step 3. Create a bucket list}} 
           
      $joinBucket \leftarrow recList + recBucketRep + recBucketUserProp$;  
      
     $joinBucket \leftarrow$ sort(joinBucket); 

     \tcp*[h]{\textcolor{blue}{Step 4. Perform selection of items with three phases}}\\

    $userCounts$,  $userGenCounts$ , $contCounts$  $\leftarrow$  $\emptyset$; 
   
   $joinBucket \leftarrow$ \textbf{selectWithHardConstraints}($joinBucket$, $recBucketRep$, $recBucketUserProp$, $userCounts$, $userGenCounts$, $contCounts$)   \tcp*[l]{\textcolor{red}{Phase 1}}
      
 $joinBucket \leftarrow$ \textbf{selectWithSoftConstraints}($joinBucket$, $recBucketRep$, $recBucketUserProp$, 2, $userCounts$, $userGenCounts$, $contCounts$)  \tcp*[l]{\textcolor{red}{Phase 2}}
      
    $joinBucket \leftarrow$ \textbf{selectWithSoftConstraints}($joinBucket$, $recBucketRep$, $recBucketUserProp$, 3, $userCounts$, $userGenCounts$, $contCounts$)  \tcp*[l]{\textcolor{red}{Phase 3}}
     
    $reRankedList \leftarrow$ chooseSelectedItems(joinBucket); 
    
    $reRankedList \leftarrow$ sort(reRankedList) \tcp*[l]{\textcolor{blue}{sort by user and rating}}
    
  \Return $reRankedList$; 
  
 }
\caption{Pseudocode of MOReGIn algorithm}
\label{code:mainalgorithm}
}
\end{algorithm}

\begin{algorithm}[] 
{ \footnotesize 
 
define \textbf{selectWithHardConstraints($joinBucket$, $recBucketRep$, $recBucketUserProp$, $userCounts$, $uGenCounts$, $contCounts$) }
 \Begin{  
    $expectedRecordsCont \leftarrow getExpectedRecordsCont(recBucketRep)$; 
    
    $expectedRecUserGen \leftarrow getRecordsUserGen(recBucketUserProp)$; 
    
    \ForEach (\tcp*[h]{\textcolor{blue}{for each record}}) {$rec \in joinBucket$} {
  
        $userGen  \leftarrow rec.user + "-" + rec.genre$;  \\ 
            \If{ $userGen \in expectedRecUserGen$ and $rec.cont \in expectedRecordsCont$}{  
            $userCounts[rec.user] \leftarrow userCounts[rec.user] +1 $; \\ 
            $uGenCounts[rec.userGen] \leftarrow uGenCounts[rec.userGen] + 1 $; \\
            $contCounts[rec.cont] \leftarrow contCounts[rec.cont] + 1 $; \\
                \If{ expectedRecUserGen[rec.userGen]  $\geq$ userGenCounts and expectedRecordsCont[rec.cont] $\geq$  contCounts and topk $\geq$ userCounts[rec.user]  }{  
                    rec.phase $\leftarrow$ 1\tcp*[l]{\textcolor{blue}{selects element in phase 1}}
                    $joinBucket.update(rec)$\tcp*[l]{\textcolor{blue}{updates the element}}
                }    
            }  
    }
     \Return joinBucket
    }
    
define \textbf{selectWithSoftConstraints($joinBucket$, $recBucketRep$, $recBucketUserProp$, $phaseMOReGIn$, $userCounts$, $userGenCounts$, $contCounts$) }
 \Begin{  
    
    $expectedRecordsCont \leftarrow getExpectedRecordsCont(recBucketRep)$; 
    
    \ForEach (\tcp*[h]{\textcolor{blue}{for each record}}) {$rec \in joinBucket$} {
                \If{ $rec.cont \in expectedRecordsCont$}{ 
       
                 \If{$phaseMOReGIn == 2$}{
                     \If{ expectedRecordsCont[rec.cont] $\geq$  contCounts and topk $\geq$ userCounts[rec.user]  }{ 
                        $userCounts[rec.user] \leftarrow userCounts[rec.user] +1 $; \\ 
                         $contCounts[rec.cont] \leftarrow contCounts[rec.cont] + 1 $; \\
                         rec.phase $\leftarrow$ 2\tcp*[l]{\textcolor{blue}{selects element in phase 2}}
                        $joinBucket.update(rec)$\tcp*[l]{\textcolor{blue}{updates the element}}
                    } 
                }
                \If{$phaseMOReGIn == 3$}{ 
                     \If{ topk $\geq$ userCounts[rec.user]  }{ 
                        $contCounts[rec.cont] \leftarrow contCounts[rec.cont] + 1 $; \\
                        rec.phase $\leftarrow$ 3\tcp*[l]{\textcolor{blue}{selects element in phase 3}}
                        $joinBucket.update(rec)$\tcp*[l]{\textcolor{blue}{updates the element}}
                        }
                 }
                }       
     }
    \Return joinBucket
}
 
\caption{Selection methods for the MOReGIn algorithm}
\label{code:support_calibration}
}
\end{algorithm}

\textsc{MOReGIn} adjusts the recommendations according to the continent of the providers and the representation of each demographic group and seeks to make a calibration at the individual level, following the propensity of each user to rate items of a given genre. 
Formally, \textsc{MOReGIn} (see Algorithm~\ref{code:mainalgorithm}) works following four main steps. \textbf{Steps 1 and 2} are devoted to compute $\mathcal{R}_c$ and $\mathcal{P}_{ug}$, considering the ratings in the training set. 
\textbf{Step 3} computes the items that were predicted as relevant for a user by the recommender system and creates a bucket list, $joinBucket$, considering each continent-genre pair, which will store the predicted items. Each bucket comes with two attributes: $\mathcal{R}_c$ and $\mathcal{P}_{ug}$.
Specifically, the recommender system returns a list of top-$n$ recommendations (where $n$ is much larger than the cut-off value $k$, so as to be able to perform a re-ranking). 
Our starting point to fill a bucket is the relevance predicted for a user $u$ and an item $i$, $\hat{r}_{ui}$. That item will be stored in the buckets associated with each genre $g \in G_i$ and each continent $c \in C_i$ (even though an item may appear in more than one bucket, it can only be recommended only once). Each element in the bucket is a record that contains the item ID and the 
$\hat{r}_{ui}$. 
We sort each bucket considering three values. We sort out $\mathcal{R}_c$ and $\mathcal{P}_{ug}$, in ascending order to ensure the inclusion in the recommendation lists of items from genres and continents that are less represented in the dataset, and we sort in descending order by rating to enhance those products that are relevant to the user. 
Finally, \textbf{Step 4} performs a three-phase re-ranking based on the generated bucket lists. Phase 1 is where we begin, and subsequent phases occur until the top-$k$ is complete.
In detail, \textbf{Phase 1} selects items starting from the least represented continents to the most represented ones in their corresponding buckets. The algorithm selects items with these conditions: (1) the percentage of items in the recommendation list for a continent is lower or equal to the representation of the continent ($\mathcal{R}_c$); (2) the percentage of items of a given genre in the top-$k$ is lower or equal than $\mathcal{P}_{ug} \cdot k$; and (3) the number of recommended items so far is lower than $k$. \textbf{Phase 2} relaxes the restrictions of phase 1 and here condition 2 is not applied. \textbf{Phase 3} selects the items that have the greater relevance for the user, until we complete the top-$k$. That is, conditions 1 and 2 are not considered.  

\section{Experimental Evaluation}\label{sec:assessment}

\subsection{Experimental Methodology}\label{sec:experimentalsetting}

In this work, we focus on well-known state-of-the-art Collaborative Filtering algorithms: \textbf{ItemKNN}~\cite{SarwarKKR01}, \textbf{UserKNN}~\cite{HerlockerKR02}, \textbf{BPRMF}~\cite{RendleFGS09}, \textbf{SVDpp}~\cite{Koren08}, and \textbf{NeuMF}~\cite{he2017neural}). We will report the results of the original recommendation algorithm (denoted as {\bf OR}). We also consider two comparison baselines: (i) a greedy calibration algorithm~\cite{ref_calibration} (denoted as {\bf CL}) with a $\lambda$ value of 0.99 (setup defined in~\cite{ref_calibration}), which post-processes the recommendation lists generated by traditional recommender systems; and (ii) a provider fairness algorithm~\cite{GomezBS22} (denoted as {\bf PF}) that considers the providers' continent provenance as a sensitive attribute, with a re-ranking approach that regulates the share of recommendations given to the items produced in a continent (visibility) and the positions in which items are ranked in the recommendation list (exposure).

To run the recommendation models, we used the \textit{Elliot} framework~\cite{ref_elliot}, which generated the recommendations for each user that fed the input of \textsc{MOReGIn}. 
As noted in Section~\ref{subsec:characterization}, the dataset was divided into two sets, one for training (80\%) and the other for testing with the most recent ratings of each user (20\%).

For each user, we generated the top-$1000$ recommendations (denoted in the paper as the top-$n$; we remind the reader that these $n=1000$ results are not shown to the users, they are only used internally by our algorithm) to then re-rank the top-$k$ (set up to 10) through the proposed \textsc{MOReGIn} algorithm. We performed a grid search of the hyper-parameters for each model in the two datasets.  For ItemKNN and UserKNN, in both datasets, we use 50 \textit{neighbors}, a cosine \textit{similarity}, and the classical \textit{implementation}. For BPRMF, SVDpp, and NeuMf we defined 10 \textit{epochs} and 10 \textit{factors} on each dataset, except NeuMF in Movies that uses 12 \textit{factors}. The \textit{batch size} is 512 for SVDpp and NeuMF and is 1 for BPRMF on both datasets. Moreover, for {BPRMF in Movies$\thicksim$Songs}, \textit{learning rate}=0.1$\thicksim$1.346, \textit{bias regularization}=0$\thicksim$1.236, \textit{user regularization}=0.01$\thicksim$1.575, \textit{positive item regularization}=0.01$\thicksim$1.376, and \textit{negative item regularization}=0.01$\thicksim$1.624; for {SVDpp in Movies$\thicksim$Songs}, 
\textit{learning rate}=0.01$\thicksim$0.001, \textit{factors regularization}=0.1$\thicksim$0.001, and \textit{bias regularization}=0.001 in both datasets; {NeuMF in Movies$\thicksim$Songs}, 
the \textit{multi-layer perceptron}=10 in both, \textit{learning rate}=0.0025 in both, and \textit{factors regularization}=0.1$\thicksim$0.001. 

\subsection{Assessment of Disparities and Mitigation}\label{sec:assessment_disparities}

\begin{figure}[t]
\centering
\subfloat[Movies $\Delta \mathcal{V}_c$ \label{fig:movies_delta_total}]{
  \includegraphics[width=0.48\textwidth]{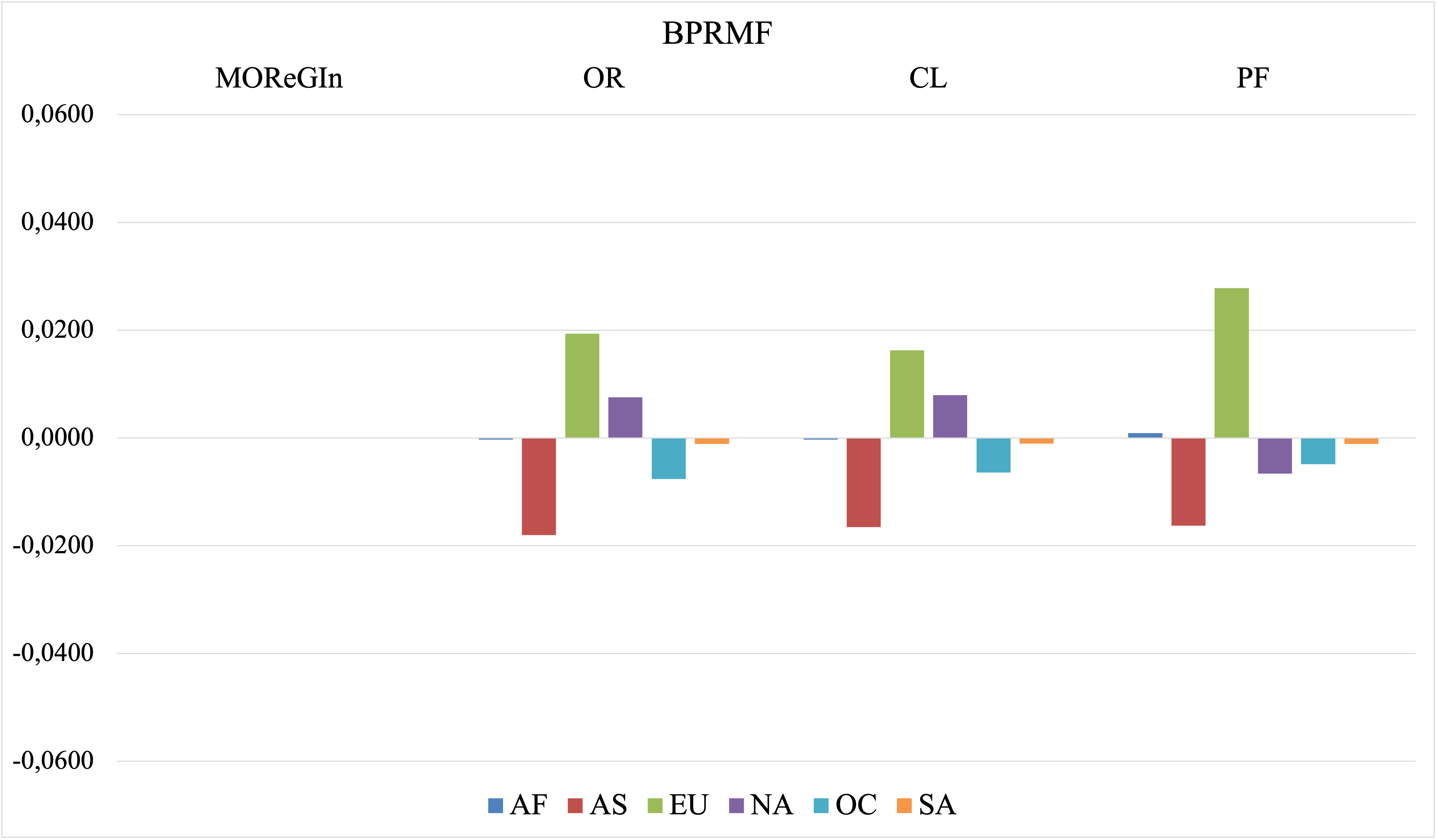}%
}\hfil
\subfloat[Movies  $\Delta \mathcal{M}_{ug}$ \label{fig:movies_delta_gen}]{
  \includegraphics[width=0.48\textwidth]{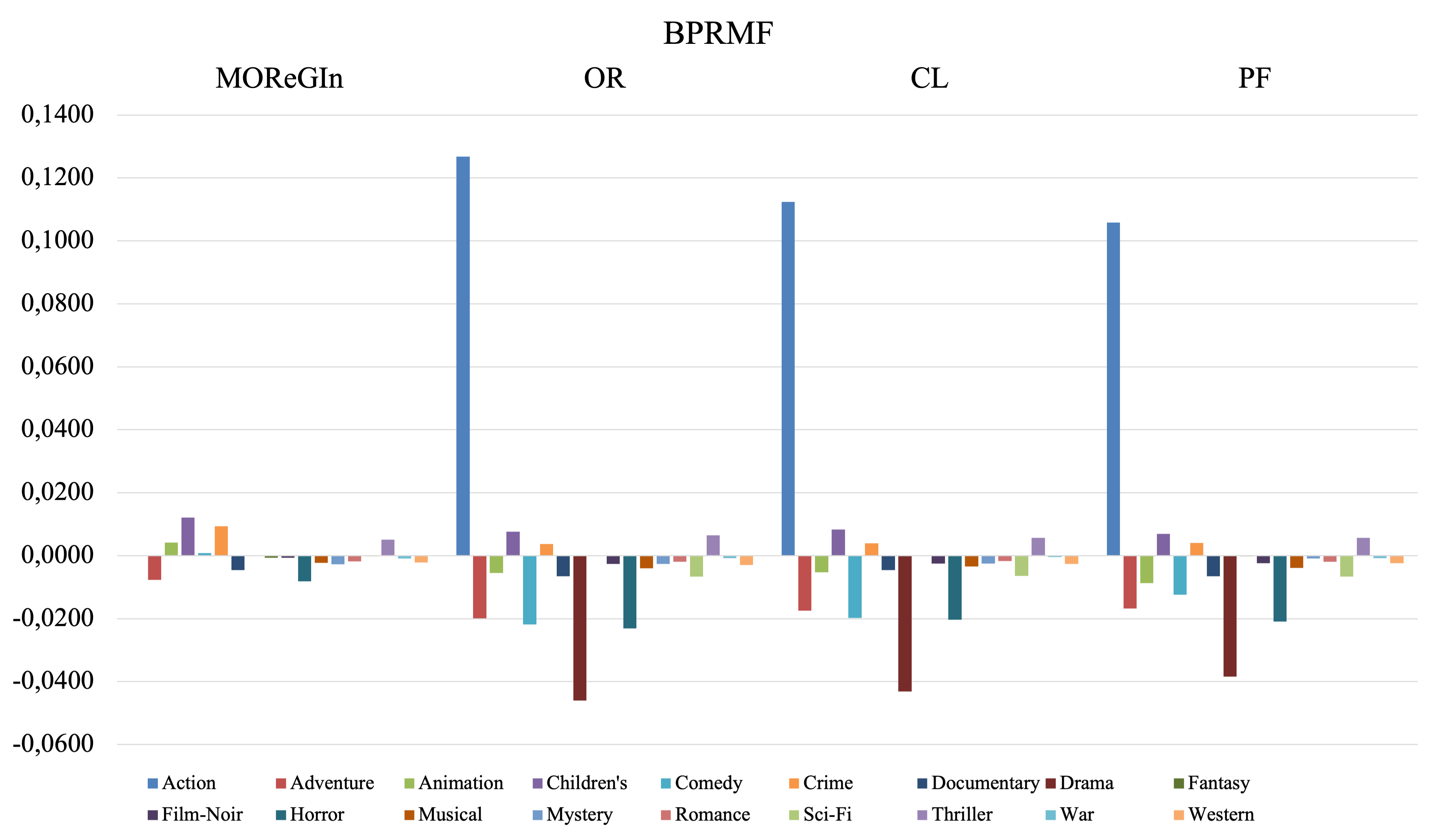}%
}
\caption{{\bf Disparity mitigation per continent (a) and miscalibration per genre (b) in BPRMF.}} 
\vspace{-0.3cm}
\label{fig:disparities_detailed}
\end{figure}

Table \ref{tab:disparities_cont} compares \textsc{MOReGIn} with the baselines in terms of the overall disparate visibility, $\Delta Total$,  for each continent. It is computed as $\forall c \in C$, $\Delta Total=\sum \Delta\mathcal{V}_c$. \textsc{MOReGIn} almost entirely reduces the disparities in both movies and song datasets, where most results are $\Delta Total=0.0000$. Although there is a little difference in the $\Delta Total$ between some approaches, these differences are more explicit, considering the provider provenance. For example, in the movie domain with the BPRMF algorithm, the $\Delta Total$ value in the OR approach is similar to that of PF. However, in a more detailed analysis of more representative continents such as NA and EU, there are notorious differences between the two approaches (i.e., 0.0075 for OR  in contrast to -0.0066 for PF in the NA continent, see the example shown in Figure~\ref{fig:movies_delta_total}).    
It is important to highlight that in both domains, our proposal mitigates the disparity regardless of the provenance of the provider, in contrast to the other algorithms that show a clear dependence on the data (i.e., the continent attribute). 

Regarding the item genres, Table~\ref{tab:disparities_gen} compares \textsc{MOReGIn} with the baselines in terms of the overall miscalibration, $\Delta Genre$,  for each continent. It is computed as $\forall g \in G$ and each user $u \in U$, $\Delta Genre=\sum \Delta\mathcal{M}_{ug}$. 
For both datasets, \textsc{MOReGIn} obtained the best $\Delta Genre$ (i.e., lowest miscalibration) in all the recommendation models. An analysis of how the algorithms behave with the different genres is shown in \figurename~\ref{fig:movies_delta_gen}. 
Although miscalibration never reaches values of $\Delta Genre$ equal to zero, our proposal always calibrates better than the baselines. 

\hlbox{Observation 2}{\textsc{MOReGIn}, by taking action on the distribution of the items per genre at the user level and provider provenance at the same time, can both calibrate and be fair to the providers. This joint effort allows us to improve the capability to calibrate the results and to be fair to providers with respect to baselines devoted solely to these purposes.}

\begin{table}[t]
\centering
\caption{{\bf Results of disparity mitigation of continents in the Movies and Songs datasets.} Each value represents the sum of disparities, $\Delta{Total}$.}
\begin{tabular}{l|cccc|cccc}
\hline
 & \multicolumn{4}{c|}{\textbf{MOVIES}} & \multicolumn{4}{c}{\textbf{SONGS}} \\
                 & \multicolumn{1}{c}{\textbf{OR}} & \multicolumn{1}{c}{\textbf{CL}} & 
                 \multicolumn{1}{c}{\textbf{PF}} & 
                 \multicolumn{1}{c|}{\textbf{MOReGIn}} &
                 \multicolumn{1}{c}{\textbf{OR}} & \multicolumn{1}{c}{\textbf{CL}} & 
                 \multicolumn{1}{c}{\textbf{PF}} & 
                 \multicolumn{1}{c}{\textbf{MOReGIn}}
                 \\\hline
\textbf{BPRMF}   & 0.0539 & \underline{0.0485} & 0.0576 & \textbf{0.0000} & 0.2637 & \underline{0.0840} & 0.2628 & \textbf{0.0000} \\
\textbf{SVDpp}   & 0.1154 & 0.1085 & \underline{0.1059} & \textbf{0.0000} & 0.2678 & \underline{0.1063} & 0.2445 & \textbf{0.0000} \\
\textbf{NeuMF}   & \underline{0.0395} & 0.0421 & 0.0638 & \textbf{0.0000} & 0.4434 & 0.4516 & \underline{0.3990} & \textbf{0.0000} \\
\textbf{UserKNN} & 0.0345 & \underline{0.0327} & 0.0328 & \textbf{0.0000} & \underline{0.0361} & 0.0575 & 0.0370 & \textbf{0.0000} \\
\textbf{ItemKNN} & 0.0431 & 0.0418 & \underline{0.0412} & \textbf{0.0000} & \underline{0.0392} & 0.0583 & 0.0420 & \textbf{0.0000} \\ \hline
\end{tabular}
\label{tab:disparities_cont}
\end{table}
\begin{table}[t]
\centering
\caption{{\bf Results of miscalibration of genres in the Movies and Songs datasets.} Each value represents the sum of miscalibrations, $\Delta{Genre}$.} 
\begin{tabular}{l|cccc|cccc}
\hline
 & \multicolumn{4}{c|}{\textbf{MOVIES}} & \multicolumn{4}{c}{\textbf{SONGS}} \\
                 & \multicolumn{1}{c}{\textbf{OR}} & \multicolumn{1}{c}{\textbf{CL}} & 
                 \multicolumn{1}{c}{\textbf{PF}} & 
                 \multicolumn{1}{c|}{\textbf{MOReGIn}} &
                 \multicolumn{1}{c}{\textbf{OR}} & \multicolumn{1}{c}{\textbf{CL}} & 
                 \multicolumn{1}{c}{\textbf{PF}} & 
                 \multicolumn{1}{c}{\textbf{MOReGIn}}
                 \\\hline
\textbf{BPRMF}   & 0.2892 & 0.2606 & \underline{0.2454} & \textbf{0.0634} & 5.5107 & \underline{0.0772} & 0.4610 & \textbf{0.0289} \\
\textbf{SVDpp}   & 0.5792 & \underline{0.5026} & 0.5694 & \textbf{0.1184} & 0.5773 & 0.1031 & \underline{0.5029} & \textbf{0.0256} \\
\textbf{NeuMF}   & 0.4596 & 0.3962 & \underline{0.3735} & \textbf{0.2901} & 1.2886 & \underline{0.7494} & 1.2202 & \textbf{0.0787} \\
\textbf{UserKNN} & 0.0743 & 0.0862 & \underline{0.0580} & \textbf{0.0392} & 0.0298 & 0.0989 & \underline{0.0291} & \textbf{0.0208} \\
\textbf{ItemKNN} & 0.2102 & 0.1966 & \underline{0.1954} & \textbf{0.0559} & 0.0890 & \underline{0.0601} & 0.0879 & \textbf{0.0205}\\ \hline
\end{tabular}
\label{tab:disparities_gen}
\end{table}

\subsection{Impact on the Quality of Recommendations}\label{sec:ndcg}

We evaluate the accuracy for the different approaches via the NDCG metric.
\begin{table}[t]
\centering
\caption{{\bf NDCG for each approach and recommendation algorithm}.}
\begin{tabular}{l|cccc|cccc}
\hline
 & \multicolumn{4}{c|}{\textbf{MOVIES}} & \multicolumn{4}{c}{\textbf{SONGS}} \\
                & \multicolumn{1}{c}{\textbf{OR}} & \multicolumn{1}{c}{\textbf{CL}} & 
                 \multicolumn{1}{c}{\textbf{PF}} & 
                 \multicolumn{1}{c|}{\textbf{MOReGIn}} &
                 \multicolumn{1}{c}{\textbf{OR}} & \multicolumn{1}{c}{\textbf{CL}} & 
                 \multicolumn{1}{c}{\textbf{PF}} & 
                 \multicolumn{1}{c}{\textbf{MOReGIn}}
                 \\\hline
\textbf{BPRMF}   & \textbf{0.3204} & 0.3144 & \underline{0.3195} & 0.3057 & 0.0034 & \textbf{0.0067} & 0.0031 & \underline{0.0055} \\
\textbf{SVDpp}   & 0.0830 & \underline{0.0888} & 0.0812 & \textbf{0.1024} & 0.0050 & \underline{0.0103} & 0.0051 & \textbf{0.0138} \\
\textbf{NeuMF}   & \underline{0.1963} & 0.1931 & 0.1956 & \textbf{0.2050} & \underline{0.0183} & 0.0098 & 0.0179 & \textbf{0.0314} \\
\textbf{UserKNN} & \underline{0.3051} & 0.2954 & 0.3030 & \textbf{0.3053} & \textbf{0.3760} & 0.1925 & \underline{0.3759} & 0.2648 \\
\textbf{ItemKNN} & \textbf{0.3229} & 0.3145 & \underline{0.3211} & 0.3131 & \textbf{0.3860} & 0.1668 & \underline{0.3857} & 0.2864 \\ \hline
\end{tabular}
\label{tab:ndcg}
\end{table}

Table \ref{tab:ndcg} shows its values for \textsc{MOReGIn} and the rest of the baselines, in all the recommendation algorithms, for Movies and Songs. 
\textsc{MOReGIn} obtained a better NDCG than the PF model, except for BPRMF and ItemKNN in the Movies dataset, and UserKNN and ItemKNN in the Songs dataset. Similar results are obtained with the CL method. 
Comparing \textsc{MOReGIn} to a non-fair approach,  MOReGIn outperforms OR models, with the exception of BPRMF and ItemKNN in the Movie domain. Except for UserKNN and ItemKNN, \textsc{MOReGIn} also outperforms the OR model in the Songs domain. 

All recommendation quality results show that the need for fairer and calibrated recommendations impacts the recommendation quality. However, beyond-accuracy perspectives, such as those offered by \textsc{MOReGIn} allows for compensating for the  minimal loss in quality with more unbiased recommendations.

\section{Conclusions and Future Work}\label{sec:conclusions}

Global and individual objectives in MORs have never been studied jointly. To study this problem, 
we provided data, by i) extending the MovieLens-1M dataset 
and ii) collecting a new dataset for song recommendation. 
The analysis of this data showed that when users rate items of a given genre, the geographic provenance of that item matters. Based on these insights, we proposed a new post-processing approach, named \textsc{MOReGIn}, that aggregates the recommended items into buckets, pairing item genres and their continent of production. 
Results show that \textsc{MOReGIn} outperforms the existing approaches at producing effective, calibrated, and provider-fair recommendations.
Future work will explore different strategies to generate recommendation lists given the generated buckets. Moreover, we will consider consumer fairness as a global perspective.

\section*{Acknowledgments}
\begin{small}
D. Contreras research was partially funded by postdoctoral project (grant No. 74200094) from ANID-Chile and by the supercomputing infrastructure of the NLHPC (ECM-02). M. Salamó was  supported  by  the  FairTransNLP-Language Project (MCIN-AEI-10.13039-501100011033-FEDER) and by the Generalitat de Catalunya (2021 SGR 00313). Maria also belongs to the Associated unit to CSIC by IIIA. 
\end{small}

\newpage

\bibliographystyle{splncs04}
\bibliography{33GomezContrerasBorattoSalamo}

\end{document}